# The Boussinesq Problem
# in Dipolar Gradient Elasticity

**H.G. Georgiadis *[1], P.A. Gourgiotis [2] and D.S. Anagnostou [1]**

[1] *Mechanics Division, National Technical University of Athens,*

*Zographou GR-15773, Greece*

[2] *Department of Mechanical and Structural Engineering, University of Trento*

*via Mesiano 77, Trento, Italy*

**Abstract** The three-dimensional axisymmetric Boussinesq problem of an isotropic half-space subjected to a concentrated normal quasi-static load is studied within the framework of linear dipolar gradient elasticity. Our main concern is to determine possible deviations from the predictions of classical linear elastostatics when a more refined theory is employed to attack the problem. Of special importance is the behavior of the new solution near to the point of application of the load where pathological singularities exist in the classical solution. The use of the theory of gradient elasticity is intended here to model the response of materials with microstructure in a manner that the classical theory cannot afford. A linear version of this theory results by considering a linear isotropic expression for the strain-energy density that depends on strain-gradient terms, in addition to the standard strain terms appearing in classical elasticity. Through this formulation, a microstructural material constant is introduced, in addition to the standard Lamé constants. The solution method is based on integral transforms and is exact. The present results show significant departure from the predictions of classical elasticity. Indeed, continuous and bounded displacements are predicted at the points of application of the concentrated load. Such a behavior of the displacement field is, of course, more natural than the singular behavior exhibited in the classical solution.

-------------------------------------------------------------------------------------
* Corresponding author. Tel.: +30 210 7721365; fax: +30 210 7721302.

  *E-mail address*: georgiad@central.ntua.gr (H.G. Georgiadis).

# 1 Introduction

The problem of determining the stress and displacement field in an isotropic half-space, which is acted upon by a concentrated point load on its surface is the celebrated Boussinesq problem (see Fig. 1). The Boussinesq solution of classical elasticity is discussed, e.g., by Love [1], Timoshenko and Goodier [2], Fung [3], Barber [4], and Selvadurai [5]. The Boussinesq solution is of fundamental importance in geotechnical applications including soil-structure interaction and stability, earthquake prediction, and rock mechanics [6,7]. In fact, the solution serves as the pertinent 3D Green's function utilized to determine the stresses and displacements induced by general load distributions, thus, rendering it a very useful tool in estimating and designing foundation settlements, provided that the elastic behavior is a reasonable assumption for the soils involved. The problem enjoys also important applications in Contact Mechanics and Tribology (see e.g. Barber [4], Gladwell [8], and Hills et al. [9]). In addition, the Boussinesq solution is used in a multitude of 3D elastostatic problems analyzed by the Boundary Element Method (see e.g. Beskos [10]).

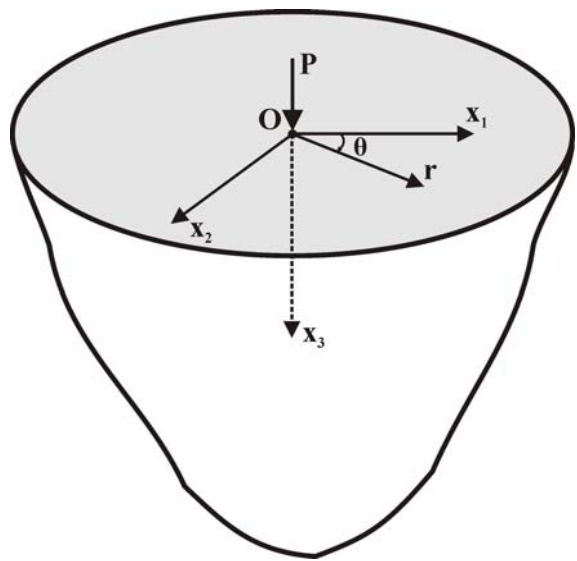

**Fig. 1** Half-space acted upon by a normal point load on its surface.

A well-known feature of the classical Boussinesq solution is that the displacement field suffers a Cauchy-type singularity at the point of application of the load. Therefore, the displacements under the point load become infinite, violating thus the basic premise of linearized elasticity. In fact, unbounded displacements occur at the point of application of the load, no matter how small the load intensity is. It is obvious, therefore, that the classical elasticity solution does not reflect (to some extent) the actual situation.

The present work aims at providing an exact solution to the three-dimensional (3D) axisymmetric Boussinesq problem that is exempt from the aforementioned deficiencies by using a more refined theory than classical elasticity. The most common version of the linear Toupin-Mindlin theory of elasticity with microstructure [11,12,13] is used, i.e. the so-called micro-homogeneous case (see Section 10 in Mindlin [12]). The general framework also appears under the names 'dipolar gradient theory' or 'strain-gradient theory' or 'grade-two theory'. A successful treatment of the 2D counterpart problem (i.e. the plane-strain Flamant-Boussinesq problem) by using the same theory was provided recently by two of the present authors (Georgiadis and Anagnostou [14]). It was found that the (exact) gradient solution predicts bounded and continuous displacements at the point of application of the load and, therefore, 'corrects' (in a boundary-layer sense) the classical Flamant-Boussinesq solution (the latter one predicts a logarithmically singular and discontinuous displacement field in the vicinity of the point of application of the concentrated line load). Finally, it is remarked that the dipolar gradient theory is a special case of the general micromorphic continuum introduced by Mindlin [12], where the micromorphic medium possesses 12 degrees of freedom: the 3 conventional translational degrees of freedom and 9 additional degrees of freedom related to the material microstructure. An interesting review and some recent applications of the micromorphic theory can be found in [15-17].

The strain-gradient theory is capable of modeling the mechanical response of solids with microstructure allowing the presence of characteristic material lengths in the governing equations. According to this, each material particle has three degrees of freedom (the displacement components – just as in the classical theories) and the micro-density does not differ from the macro-density. Also, among the three forms of that version, we choose form II of Mindlin [12] which assumes a strain-energy density that is a function of the strain tensor and its gradient. In a way, this form is a *first-step* extension of classical elasticity. We notice that the gradient of strain comprises both rotation and stretch gradients. Therefore, this version of the gradient theory is different from the standard couple-stress theory (Cosserat theory with constrained rotations) assuming a strain-energy density that depends upon the strain and the gradient of rotation only. Moreover, the dipolar gradient theory is different from the Cosserat (or micropolar) theory that takes material particles with six independent degrees of freedom (three displacement components and three rotation components, the latter involving rotation of a micro-medium w.r.t. its surrounding medium). A comprehensive review article on generalized continuum theories has recently been written by Maugin [18].

The Toupin-Mindlin gradient theory had already some successful applications on stress concentration elasticity problems concerning holes and inclusions, during the 1960s and the 1970s (see e.g. Cook and Weitsman [19], Eshel and Rosenfeld [20,21]). More recently, this approach and related extensions have been employed to analyze various problems involving, among other areas, fracture (Chen et al. [22], Shi et al. [23], Georgiadis [24], Radi and Gei [25], Grentzelou and Georgiadis [26,27], Gourgiotis and Georgiadis [28], Aravas and Giannakopoulos [29], Aslan and Forest [30], Gourgiotis et al. [31], Piccolroaz et al. [32], Sciarra and Vidoli [33]), wave propagation (see e.g. Maugin and Miled [34], Engelbrecht et al. [35], dell'Isola et al. [36], Polyzos and Fotiadis [37], Gourgiotis et al. [38]), poroelasticity (Madeo et al. [39]), mechanics of defects (Lazar and Maugin [40,41], Forest [42]), and stress concentration due to discrete loadings (Georgiadis and Anagnostou [14], Exadaktylos [43], Lazar and Maugin [44]).

Appropriate length scales for strain gradient elasticity theories were discussed by, among others, Chen et al. [22], Georgiadis et al. [13], and Polyzos and Fotiadis [37]. For instance, Chen et al. [22] developed a continuum model for cellular materials and found that the continuum description of these materials obey an elasticity theory with strain-gradient effects. In the latter study, the intrinsic material length was naturally identified with the cell size. Another example of successful modeling of microstructure and size effects in elastically deformed solids includes the propagation of waves of small wavelengths. Indeed, in electronic-device applications, surface-wave frequencies on the order of GHz are often used and therefore wavelengths on the micron order appear (see e.g. White [45], and Farnell [46]). In such situations, *dispersion phenomena* of Rayleigh waves at high frequencies can only be explained on the basis of a gradient elasticity theory (Georgiadis et al. [13]). In addition, the latter study provides an estimate for a microstructural parameter (i.e. the so-called gradient coefficient $c$) employed in some simple material models. This was effected by considering that the material is composed wholly of unit cells having the form of cubes with edges of size $2h$ and comparing the forms of dispersion curves of Rayleigh waves obtained by the Toupin-Mindlin approach with the ones obtained by the atomic-lattice analysis. It was found that $c$ is of the order of $(0.1h)^2$. Generally, theories with elastic strain gradient effects are intended to model situations where the intrinsic material lengths are of the order of 0.1 – 10 microns [23]. Since the strengthening effects arising from strain gradients become important when these gradients are large enough, these effects will be significant when the material is deformed in *very small* volumes, such as in the immediate vicinity of crack tips, notches, small holes and inclusions, and micrometer indentations.

Now, we concentrate on the subject of the present work, i.e. the 3D axisymmetric Boussinesq problem within the context of dipolar gradient elasticity. In the literature, there are only two results closely related to this subject. The first one is due to Karlis et al. [47]. This is a numerical study employing the Boundary Element Method. In this work, results only for the vertical surface displacement were obtained showing a bounded behavior at the point of application of the concentrated load. However, no results were presented for the behavior of the radial surface displacement or the influence of the material constants on the solution. The second result is due to Gao and Zhou [48], who evaluated the surface displacements in the 3D Boussinesq problem by using the method of Hankel transforms. However, they did not provide analytical expressions for the vertical and radial surface displacements in gradient elasticity since the inversions of the pertinent integral transforms were performed only numerically. In addition, no results were presented regarding the influence of the Poisson's ratio in the gradient solution. Moreover, the important issue of uniqueness of the solution in gradient elasticity was not discussed. It should be noticed that in the present analytical study all these matters are addressed in detail.

Other efforts set out to alleviate the field singularities in the 3D Boussinesq problem by the use of different generalized continuum theories are mentioned now. In the context of micropolar (Cosserat) theory, the Boussinesq problem was analyzed by Dhaliwal [49], Sandru [50], Khan and Dhaliwal [51], and Dyszlewicz [52]. It is noticed that in micropolar elasticity solutions, displacement and stress singularities occur again. The orders of these singularities are the same as in classical theory, but the structure of the fields is different. On the other hand, Nowinski [53] utilizing the Kröner-Eringen theory of non-local elasticity showed that all stress components remain finite at the point of application of the concentrated load. However, although this result is more realistic than the respective result of the classical theory, still the magnitude of stresses predicted by the non-local theory is extremely high. Finally, Simmonds and Warne [54] examined the nonlinear elastic Boussinesq problem within the framework of classical *finite elasticity*. They obtained asymptotic solutions in the case of only a *tensile* point load for an incompressible generalized neo-Hookean (power-law) material and a compressible Blatz-Ko material. It was shown that the former solid, if sufficiently stiffer than the conventional neo-Hookean material, may support a finite displacement under the point load, but the compressible Blatz-Ko solid does not support a finite displacement.

Our solution is based on integral transforms and is *exact*. Certainly, such an analytical solution has an advantage over numerical solutions. A simple but yet rigorous version of the Toupin-Mindlin theory of dipolar gradient elasticity involving only one additional material constant (the so-called gradient coefficient) is employed that leads to a solution not exhibiting the singular behavior described above. It is worth noting that, contrary to the previous numerical studies by Karlis et al. [47] and Gao and Zhou [48], we provide here *explicitly* the terms in the gradient solution that eliminate the Cauchy type singularities of classical elasticity. In this sense, the new solution predicts therefore a more natural material response. In particular, the vertical surface displacement attains a constant (bounded) value at the point of application of the load which depends upon the load intensity and Poisson's ratio. An estimate of this value is provided explicitly in terms of the Poisson's ratio and the characteristic material length. Also, the radial displacement becomes zero at the same point – this is justified by the axisymmetric character of the Boussinesq problem. It is shown that the occurrence of bounded displacements at the point of application of the concentrated load implies that the total strain energy in a small region surrounding the singular point vanishes, as the size of the region tends to zero. Consequently, an analogue of Kirchhoff's theorem guarantees *uniqueness* of solution for the concentrated load problem in gradient elasticity [26]. In addition, we remark that the bounded displacement field depends now upon the load intensity allowing, within the new context, a sound meaning of the load-carrying capacity of the medium. This may have important implications for more general contact problems and the Boundary Element Method.

In closing, we note that the occurrence of bounded fields in singular stress concentration problems is not uncommon in the context of dipolar gradient elasticity. For instance, in the 2D plane-strain Flamant-Boussinesq and Kelvin problems bounded and continuous displacements were predicted at the point of application of the loads (Georgiadis and Anagnostou [14], Exadaktylos [43], Lazar and Maugin [44]). In dislocation / disclination problems, gradient elasticity predicts finite strain and stress fields at the defect line (see e.g. Lazar and Maugin [40,41]). In addition, in 2D crack problems (see e.g. Gourgiotis and Georgiadis [28]), the near tip strain field was found to be bounded in the gradient solution exhibiting an $r^{1/2}$-variation ($r$ is the distance from the crack tip) contrary to the singular $r^{-1/2}$-variation of classical elasticity. The latter observation concurs with the uniqueness theorem for crack problems in gradient elasticity, where the necessary conditions for uniqueness are bounded displacements and strains around the crack tip (Grentzelou and Georgiadis [26]). All these results have an explanation because, in general, materials governed by dipolar gradient theory tend to behave in a more rigid way (having increased stiffness) as compared to materials governed by classical continuum theories.

## 2  Basic equations of dipolar gradient elasticity

In this Section, we will give a brief account of form II of Mindlin's theory of dipolar gradient elasticity. More detailed presentations can be found in some recent papers (Georgiadis et al. [13], Gourgiotis et al. [31,38], Georgiadis and Grentzelou [55]) and in the fundamental papers by Toupin [11], Mindlin [12], Bleustein [56], and Mindlin and Eshel [57].

One may start from the following expression of the elastic strain-energy density

$$W \equiv W\left(\varepsilon_{pq}, \partial_r \varepsilon_{pq}\right) , \qquad (1)$$

which is assumed to be a positive definite function and depends upon *both* the strain and its gradient. In the above equation, a Cartesian rectangular coordinate system $Ox_1 x_2 x_3$ is considered for a 3D continuum (indicial notation and the summation convention will be used throughout), $\partial_p(\ ) \equiv \partial(\ )/\partial x_p$, the Latin indices span the range (1,2,3), $\varepsilon_{pq} = (1/2)\left(\partial_p u_q + \partial_q u_p\right) = \varepsilon_{qp}$ is the linear strain tensor, and $u_q$ is the displacement vector. Small strains and displacements are assumed, and although (1) allows for non-linear constitutive behavior, we will confine attention here to a linear constitutive law. The strain gradient $\partial_r \varepsilon_{pq}$ has 18 independent components and comprises *both* rotation and stretch gradients [57].

Further, stresses can be defined in the standard variational manner

$$\tau_{pq} \equiv \frac{\partial W}{\partial \varepsilon_{pq}} , \qquad m_{rpq} \equiv \frac{\partial W}{\partial\left(\partial_r \varepsilon_{pq}\right)} , \qquad (2a,b)$$

where $\tau_{pq}$ is the monopolar stress tensor expressed in dimensions of [force][length]$^{-2}$, $m_{rpq}$ is the dipolar (or double) stress tensor (a third-rank tensor) expressed in dimensions of [force][length]$^{-1}$, and the following symmetries for the monopolar and dipolar stress tensors are noticed: $\tau_{pq} = \tau_{qp}$ and $m_{rpq} = m_{rqp}$. The dipolar stress tensor follows from the notion of dipolar forces, which are anti-parallel forces acting between the micro-media contained in the continuum with microstructure. As explained by Green and Rivlin [58] and Jaunzemis [59], the notion of multipolar forces arises from a series expansion of the mechanical power containing higher-order velocity gradients. A physical interpretation of these multipolar forces employing discrete structural models has been given recently by Alibert et al. [60].

Then, the equations of equilibrium (global equilibrium) and the traction boundary conditions along a boundary (local equilibrium) can be obtained from variational considerations [12,56]. Assuming the absence of body forces, the appropriate expression of the Principle of Virtual Work is written as [56]

$$\int_V \left[ \tau_{pq} \delta \varepsilon_{pq} + m_{rpq} \delta(\partial_r \varepsilon_{pq}) \right] dV = \int_S t_q^{(n)} \delta u_q \, dS + \int_S T_{qr}^{(n)} \partial_q (\delta u_r) \, dS \quad , \tag{3}$$

where the symbol $\delta$ denotes weak variations and it acts on the quantity existing on its right. In the above equation, $t_q^{(n)}$ is the *true* force surface traction, $T_{pq}^{(n)}$ is the *true* double force surface traction, and $n_p$ is the outward unit normal to the boundary along a section inside the body or along the surface of it. Examples of the latter tractions along the surface of a 2D half-space are given in Fig. 2.

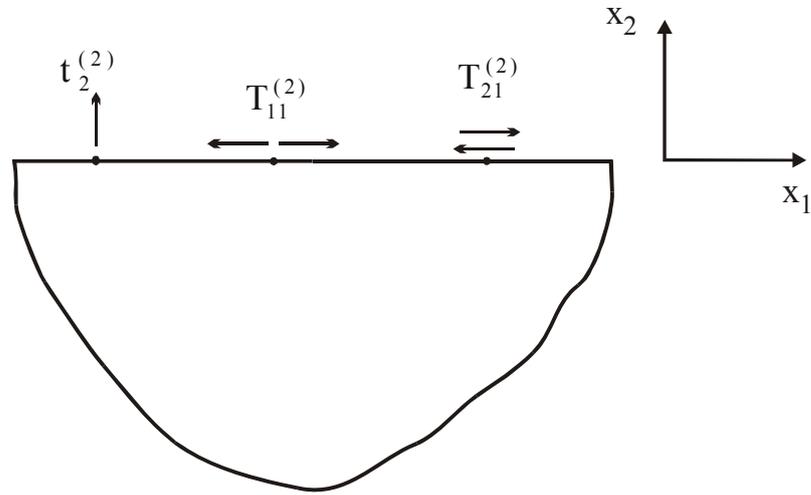

**Fig. 2** Positively oriented true monopolar and dipolar tractions on the surface of a half-space.

The equations of equilibrium and the traction boundary conditions take the following form

$$\partial_p(\tau_{pq} - \partial_r m_{rpq}) = 0 \quad \text{in} \quad V, \tag{4}$$

$$P_q^{(n)} = n_p(\tau_{pq} - \partial_r m_{rpq}) - D_p(n_r m_{rpq}) + (D_j n_j) n_r n_p m_{rpq} \quad \text{on} \quad bdy, \tag{5}$$

$$R_q^{(n)} = n_r n_p m_{rpq} \quad \text{on} \quad bdy, \tag{6}$$

$$E_q = \left[ n_r k_p m_{rpq} \right] \quad \text{on} \quad edge\ C, \tag{7}$$

where $V$ is the region (open set) occupied by the body, $bdy$ denotes any boundary along a section inside the body or along the surface of it, $D_p(\ ) \equiv \partial_p(\ ) - n_p D(\ )$ is the surface gradient operator, $D(\ ) \equiv n_p \partial_p(\ )$ is the normal gradient operator. Further, $C$ denotes every edge formed by the intersection of two portions, say $S_1$ and $S_2$ of the (closed) bounding surface $S$, and the double brackets [ ] indicate that the enclosed quantity is the difference between the values on $S_1$ and $S_2$. Also, the vector **k** is defined as $k_q = e_{rpq} s_r n_p$, where $s_q$ is the unit tangent vector to the curve $C$, and $e_{qkl}$ is the Levi-Civita permutation symbol. Furthermore, according to Bleustein [56], $P_q^{(n)} \equiv t_q^{(n)} + (D_r n_r) n_p T_{pq}^{(n)} - D_p T_{pq}^{(n)}$ is the auxiliary force traction, $R_q^{(n)} \equiv n_p T_{pq}^{(n)}$ is the auxiliary double force traction and $E_q \equiv \left[ k_p T_{pq} \right]$ is a line load defined on the edge $C$.

At this point, it is worth noting that dell'Isola and Seppecher [61] and, dell'Isola et al. [62], employing the principle of virtual power, derived formally explicit representation formulas for grade-$N$ continua, expressing the tractions (contact interactions) as functions of the shape of an arbitrary section of the body (Cauchy cut).

Finally, pertinent *kinematical* boundary conditions were derived by Georgiadis and Grentzelou [55] in the context of the Principle of Complementary Virtual Work for dipolar theory. However, these conditions are omitted here since they are not relevant to our specific problem.

Introducing the constitutive equations utilized here is now in order. The simplest possible linear and isotropic equations of gradient elasticity result from the following strain-energy density function [13,41]

$$W = (1/2)\lambda \varepsilon_{pp} \varepsilon_{qq} + \mu \varepsilon_{pq} \varepsilon_{pq} + (1/2)\lambda c (\partial_r \varepsilon_{pp})(\partial_r \varepsilon_{qq}) + \mu c (\partial_r \varepsilon_{pq})(\partial_r \varepsilon_{pq}), \tag{8}$$

where $c$ is the gradient coefficient having dimensions of [length]$^2$, and $(\lambda, \mu)$ are the standard Lamé constants with dimensions of [force][length]$^{-2}$. In this way, only one new material constant is introduced with respect to classical linear isotropic elasticity. Combining (2) with (8) provides the following constitutive equations

$$\tau_{pq} = \lambda \delta_{pq} \varepsilon_{jj} + 2\mu \varepsilon_{pq} , \quad m_{rpq} = c\partial_r \left(\lambda \delta_{pq} \varepsilon_{jj} + 2\mu \varepsilon_{pq}\right) = c\partial_r \tau_{pq} , \qquad (9a,b)$$

where $\delta_{pq}$ is the Kronecker delta.

As Lazar and Maugin [41] pointed out, the particular choice of (8) is physically justified and possesses a notable symmetry. To expose this symmetry, we first consider the general expression (definition) of the strain-energy density $W \equiv \int_0^{\varepsilon_{pq}} \tau_{pq} d\varepsilon_{pq} + \int_0^{\partial_r \varepsilon_{pq}} m_{rpq} d(\partial_r \varepsilon_{pq})$, which for a linear constitutive law takes the form $W = (1/2)\tau_{pq}\varepsilon_{pq} + (1/2)m_{rpq}\partial_r \varepsilon_{pq}$. Then, by virtue of (9b) the strain-energy density in (8) takes the form $W = (1/2)\tau_{pq}\varepsilon_{pq} + c(1/2)(\partial_r \tau_{pq})(\partial_r \varepsilon_{pq})$. Hence, by using the special choice (8) for the strain energy density, we have obtained a simple but still rigorous version of the Toupin-Mindlin gradient elasticity theory which is a simple strain gradient as well as stress gradient theory. In particular, the elastic energy is symmetric with respect to the strain and the stress, and also with respect to the strain gradient and the stress gradient. Notice that fully anisotropic constitutive relations have been used in deriving general results (energy theorems, uniqueness, balance laws and energy release rates) in recent works on gradient elasticity (Grentzelou and Georgiadis [26,27], Georgiadis and Grentzelou [55]), but use of the general relations poses serious difficulties in solving specific boundary value problems. Therefore, the assumption of isotropy and the simplification using a *single* material length mentioned above greatly facilitates the analysis of boundary value problems of gradient elasticity. The full constitutive relations in the isotropic case involve five material constants besides the two Lamé constants (Mindlin [12], dell'Isola et al. [63]). It is worth noting that recently two of the present authors investigated the 2D Flamant-Boussinesq problem in an isotropic half-space using the *exact* Toupin-Mindlin theory of gradient elasticity (i.e. with 5 additional material constants). The results showed bounded behavior for the displacement field and also the existence of many types of boundary layers in the vicinity of the concentrated load depending on the microstructural material parameters (Gourgiotis et al. [64]).

Further, it is noticed that *uniqueness* theorems have been proved on the basis of positive definiteness of the strain-energy density in cases of regular and singular (crack problems) fields in the recent works of Grentzelou and Georgiadis [26], and Georgiadis and Grentzelou [55], respectively. As shown by Georgiadis et al. [13], the restriction of positive definiteness of $W$ requires the following inequalities for the material constants appearing in the theory employed here: $(3\lambda+2\mu)>0$, $\mu>0$, $c>0$. In addition, *stability* for the field equations in the general inertial case was proved in [13] and to accomplish this, the condition $c>0$ is a necessary one.

In summary, Equations (4)-(7) and (9) are the governing equations for the isotropic linear dipolar gradient elasticity. In particular, in view of (9b) the equilibrium equations (4) take the following form

$$\partial_p\left(\tau_{pq}-c\partial_r\partial_r\tau_{pq}\right)=\left(1-c\nabla^2\right)\partial_p\tau_{pq}=0 ,\qquad(10a)$$

which, by taking into account the constitutive equation (9a) and the definition of the linear strain tensor, can be written in terms of the displacements as

$$\left(1-c\nabla^2\right)\left[\nabla^2\mathbf{u}+\left(1-2\nu\right)^{-1}\nabla(\nabla\cdot\mathbf{u})\right]=0 ,\qquad(10b)$$

where $\nabla^2(\ )$ is the Laplace operator, $\nu=\lambda/[2(\lambda+\mu)]$ is the Poisson's ratio, and the absence of body forces and couples is assumed. In the limit $c\to 0$, the Navier-Cauchy equations of classical linear isotropic elasticity are recovered from (10b). Clearly, Eqs. (10a) and (10b) reveal the *singular-perturbation* character of the gradient theory. Therefore, *boundary-layer* effects are to be expected in solutions to specific problems.

Finally, applying the gradient and the curl operator to Eq. (10b), we obtain the following relations for the dilatation and the rotation, respectively

$$\left(1-c^2\nabla^2\right)\nabla^2 e=0 , \quad \left(1-c^2\nabla^2\right)\nabla^2\boldsymbol{\omega}=0 ,\qquad(11a,b)$$

where $e\equiv\nabla\cdot\mathbf{u}$ is the dilatation (volumetric strain), and $\boldsymbol{\omega}\equiv(1/2)\nabla\times\mathbf{u}$ is the rotation vector. In indicial notation, the rotation vector is written as $\omega_q=(1/2)e_{qkl}\,\partial_k u_l$, whereas the rotation tensor is written as $\omega_{pq}=(1/2)(\partial_p u_q-\partial_q u_p)$.

## 3 Formulation and transformed solution for the axisymmetric Boussinesq problem

Consider a homogeneous isotropic body occupying the half-space $x_3 \geq 0$. The body is acted upon by a normal point load of intensity $P$ at a point on its free surface ($x_3 = 0$). This point is taken as the origin of a Cartesian rectangular coordinate system (see Fig. 1). We have also introduced a cylindrical polar coordinate system, which will be used in the subsequent analysis.

The traction boundary conditions along the surface $x_3 = 0$ with $\mathbf{n} = (0,0,-1)$ follow from (5) and (6) and are written as

$$P_1^{(n)} = (\tau_{31} - \partial_1 m_{131} - \partial_2 m_{231} - \partial_3 m_{331}) - \partial_1 m_{311} - \partial_2 m_{321} = 0 \ , \tag{12a}$$

$$P_2^{(n)} = (\tau_{32} - \partial_1 m_{132} - \partial_2 m_{232} - \partial_3 m_{332}) - \partial_1 m_{312} - \partial_2 m_{322} = 0 \ , \tag{12b}$$

$$P_3^{(n)} = (\tau_{33} - \partial_1 m_{133} - \partial_2 m_{233} - \partial_3 m_{333}) - \partial_1 m_{313} - \partial_2 m_{323} = -P\delta(x_1)\delta(x_2) \ , \tag{12c}$$

$$R_1^{(n)} = m_{331} = 0 \ , \tag{13a}$$

$$R_2^{(n)} = m_{332} = 0 \ , \tag{13b}$$

$$R_3^{(n)} = m_{333} = 0 \ . \tag{13c}$$

where $\delta(\ )$ is the Dirac delta distribution. It appears above with the standard 'symbolic' sense.

Although the existing *axisymmetry* (i.e. circular symmetry) w.r.t. the $x_3$-axis suggests the use of the Hankel transform, we chose here, instead, to suppress the dependence of the problem on the space-variables $(x_1, x_2)$ through the use of the double two-sided (or bilateral) Laplace transform (see e.g. van der Pol and Bremmer [65], and Carrier et al. [66]). This is because we intend to use the present basic integral-transform analysis for more general *non-axisymmetric* situations like the Cerruti problem in dipolar gradient elasticity (the latter problem will be treated in a subsequent paper). Certainly, the fact that we deal with an axisymmetric field in our specific problem will emerge in the course of solving the problem. For some applications of the double two-sided Laplace transform in elasticity problems, we refer to Brock and Rodgers [67], and Georgiadis and Lykotrafitis [68].

The double bilateral transform pair (direct and inverse operation) is defined as

$$f^*(p,q,x_3) = \int_{-\infty}^{\infty}\int_{-\infty}^{\infty} f(x_1,x_2,x_3)\cdot e^{-px_1-qx_2} dx_1\, dx_2 \quad, \tag{14a}$$

$$f(x_1,x_2,x_3) = \left(\frac{1}{2\pi i}\right)^2 \int_{\Gamma_1}\int_{\Gamma_2} f^*(p,q,x_3)\cdot e^{px_1+qx_2} dp\, dq \quad, \tag{14b}$$

where the direct transform suppresses the space-variables $(x_1, x_2)$. In what follows, the double bilateral direct transform is denoted by an asterisk. It is also noticed that the variables $p$ and $q$ should be treated as *complex* and the integration paths $\Gamma_1$ and $\Gamma_2$ are lines parallel to the imaginary axis in the associated transform plane and lie *within* the region of analyticity. Transforming (10b) with (14a) gives a system of ODEs for $(u_1^*, u_2^*, u_3^*)$ written in the following compact form (for details see Appendix A)

$$[K]\begin{bmatrix} u_1^* \\ u_2^* \\ u_3^* \end{bmatrix} \equiv \begin{bmatrix} K_{11} & K_{12} & K_{13} \\ K_{21} & K_{22} & K_{23} \\ K_{31} & K_{32} & K_{33} \end{bmatrix}\begin{bmatrix} u_1^* \\ u_2^* \\ u_3^* \end{bmatrix} = \begin{bmatrix} 0 \\ 0 \\ 0 \end{bmatrix} \quad, \tag{15}$$

where the components $K_{ij}$ $(i,j=1,2,3)$ of the *symmetric* differential operator $[K]$ are defined as

$$K_{11} = \left[(1-2\nu)(d^2+q^2) + 2(1-\nu)p^2\right]\left[1 - c(d^2+p^2+q^2)\right] \quad, \tag{16a}$$

$$K_{12} = K_{21} = pq\left[1 - c(d^2+p^2+q^2)\right] \quad, \tag{16b}$$

$$K_{13} = K_{31} = pd\left[1 - c(d^2+p^2+q^2)\right] \quad, \tag{16c}$$

$$K_{22} = \left[(1-2\nu)(d^2+p^2) + 2(1-\nu)q^2\right]\left[1 - c(d^2+p^2+q^2)\right] \quad, \tag{16d}$$

$$K_{23} = K_{32} = qd\left[1 - c(d^2+p^2+q^2)\right] \quad, \tag{16e}$$

$$K_{33} = \left[(1-2\nu)(p^2+q^2) + 2(1-\nu)d^2\right]\left[1 - c(d^2+p^2+q^2)\right] \quad, \tag{16f}$$

with $d^n(\ ) \equiv d^n(\ )/dx_3^n$. It is noted that Eqs. (15) form a homogeneous system of ODEs with constant coefficients, the solution can then be obtained by utilizing a *symbolic* method where the differential operator is treated as parameter (see e.g. Sneddon [69]). Accordingly, the set of homogeneous differential equations in (15) has a solution different than the trivial one if and only if the determinant of $[K]$ is zero. Hence,

$$\det[K] = \left(d^2 + p^2 + q^2\right)^3 \left[1 - c\left(d^2 + p^2 + q^2\right)\right]^3 = 0 \;, \tag{17}$$

which has two triple roots: $d = \pm\left(-p^2 - q^2\right)^{1/2}$ and $d = \pm\left(1/c - p^2 - q^2\right)^{1/2}$. The former are the same as in classical elasticity, whereas the latter reflect the presence of gradient effects. The general solution of (15) is obtained after some rather extensive algebra and it has the following form that is bounded as $x_3 \to +\infty$

$$u_1^*(p,q,x_3) = p^{-1}\left(A_1\beta - A_2 q - A_3\left[3 - 4\nu + \frac{p^2 x_3}{\beta}\right]\right)e^{-\beta x_3} + B_1 e^{-\gamma x_3} \;, \tag{18}$$

$$u_2^*(p,q,x_3) = \left(A_2 - A_3\frac{q x_3}{\beta}\right)e^{-\beta x_3} + B_2 e^{-\gamma x_3} \;, \tag{19}$$

$$u_3^*(p,q,x_3) = \left(A_1 + A_3 x_3\right)e^{-\beta x_3} + B_3 e^{-\gamma x_3} \;, \tag{20}$$

where the quantities $\left(A_j, B_j\right)$ with $(j = 1, 2, 3)$ are yet unknown functions of $(p, q)$ that will be determined through enforcing the boundary conditions of the problem. Also, the following definitions are employed in (18)-(20):

$$\beta \equiv \beta(p,q) = \left(-p^2 - q^2\right)^{1/2} \;, \quad \gamma \equiv \gamma(p,q) = \left(a^2 - p^2 - q^2\right)^{1/2} \;, \tag{21}$$

with $a = (1/c)^{1/2}$.

Further, a new complex variable is defined through $\zeta^2 = p^2 + q^2$, allowing the placement of necessary branch cuts in the $\zeta$-plane for the functions $\beta \equiv \beta(\zeta) = \left(\varepsilon^2 - \zeta^2\right)^{1/2}$ (with $\varepsilon$ being a real number such that $\varepsilon \to +0$) and $\gamma \equiv \gamma(\zeta) = \left(a^2 - \zeta^2\right)^{1/2}$. In fact, introducing $\varepsilon$ facilitates the introduction of branch cuts for the complex function $\beta = \left(-\zeta^2\right)^{1/2}$ (see e.g. Carrier et al. [66], and Georgiadis [24] for this convenient way of defining branch cuts).

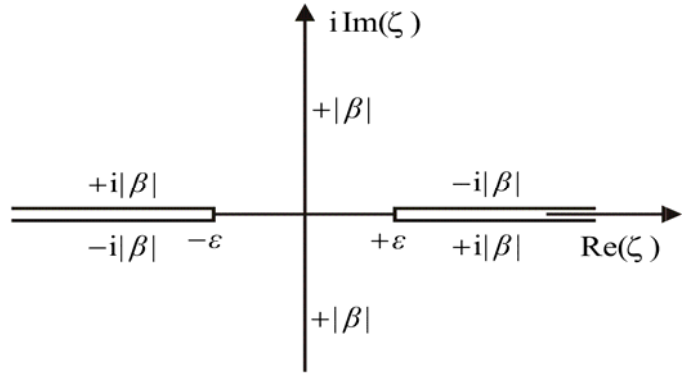

**Fig. 3** Branch cuts for the function $\beta(\zeta)$.

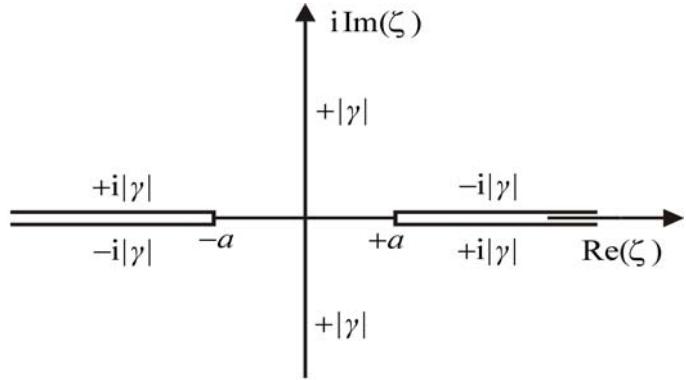

**Fig. 4** Branch cuts for the function $\gamma(\zeta)$.

It is emphasized that in order to obtain a *bounded* solution as $x_3 \to +\infty$, the $\zeta$-plane should be cut in the way shown in Figs. 3 and 4. This introduction of branch cuts secures that the functions $(\beta,\gamma)$ are single-valued and that $\text{Re}(\beta) \geq 0$ and $\text{Re}(\gamma) \geq 0$ in the cut-plane.

The transformed general expressions for the stresses and dipolar stresses that enter the boundary conditions are quoted in Appendix B. Transforming the boundary conditions (12)-(13) with (14a) and enforcing (B1)-(B7) provides a $6 \times 6$ system of algebraic equations for the unknown functions $(A_j, B_j)$, with $(j = 1, 2, 3)$. After some algebra involving manipulations by hand and the symbolic program MATHEMATICA$^{\text{TM}}$, a solution is obtained and is given in Appendix C. Finally, the transformed displacements are found to be

$$u_1^*(p,q,x_3) = -\frac{Pp}{2\mu\beta^2 N}\left[4\nu c^2\beta^3\gamma + (1-2\nu)(2c^2\beta^2\gamma^2 - \nu + 1)\right]e^{-\beta x_3}$$

$$+ \frac{Pp}{2\mu\beta N}\left[1-\nu - 2c^2\beta^2\gamma(\beta-\gamma)\right]x_3\, e^{-\beta x_3} + \frac{Pcp}{\mu N}\left[c\gamma^2 - \nu\right]e^{-\gamma x_3}, \quad (22)$$

$$u_2^*(p,q,x_3) = -\frac{Pq}{2\mu\beta^2 N}\left[4\nu c^2\beta^3\gamma + (1-2\nu)(2c^2\beta^2\gamma^2 - \nu + 1)\right]e^{-\beta x_3}$$

$$- \frac{Pq}{2\mu\beta N}\left[2c^2\beta^2\gamma(\beta-\gamma) - 1 + \nu\right]x_3\, e^{-\beta x_3} + \frac{Pcq}{\mu N}\left[c\gamma^2 - \nu\right]e^{-\gamma x_3}, \quad (23)$$

$$u_3^*(p,q,x_3) = \frac{P}{\mu\beta N}\left[(3-2\nu)c^2\beta^3\gamma - (1-\nu)(2c^2\beta^2\gamma^2 - \nu + 1)\right]e^{-\beta x_3}$$

$$+ \frac{P}{2\mu N}\left[2c^2\beta^2\gamma(\beta-\gamma) - 1 + \nu\right]x_3\, e^{-\beta x_3} - \frac{Pc\beta^2}{\mu\gamma N}\left[c\beta^2 + \nu\right]e^{-\gamma x_3}, \quad (24)$$

where

$$N = 4(c\beta\gamma)^3 - (1+2c\beta^2)\left(2(c\beta\gamma)^2 - \nu + 1\right). \quad (25)$$

## 4 Laplace transform inversion

In what follows, we focus attention only on the surface displacements ($x_3 = 0$) with a view towards obtaining an *explicit* analytical solution. Such a solution is intended to determine the behavior of the displacement field *near* to the point of application of the concentrated load and will allow detecting possible deviations from the predictions of classical theory. Certainly, this is our main concern here. Notice, however, that determining the field at points 'inside' the half-space ($x_3 \neq 0$) may follow along the same general lines of the present analysis but it will involve much additional numerical work because of the presence of the terms $\exp(-\beta x_3)$ and $\exp(-\gamma x_3)$ in the transformed displacements.

In view of the definition (14b) and the particular choice of branch cuts (Figs. 3 and 4), one can write the surface displacements in the form

$$u_j(x_1,x_2,x_3=0) = \left(\frac{1}{2\pi i}\right)^2 \int_{-i\infty}^{+i\infty}\int_{-i\infty}^{+i\infty} u_j^*(p,q,x_3=0)\cdot e^{px_1+qx_2}\,dp\,dq\ , \quad (j=1,2,3) \tag{26}$$

where the transformed displacements $u_j^*$ are given in (22)-(24). Next, axisymmetry of the problem will become clear and be exploited. To this end, we set $p=-i\xi$, $q=-i\eta$, where $(\xi,\eta)\in\Re$, and consider the polar coordinates $(r,\theta)$ and $(\rho,\varphi)$ defined through the relations: $x_1+ix_2=re^{i\theta}$ and $\xi+i\eta=\rho e^{i\varphi}$. The first set of polar coordinates refers to the physical plane $(x_1,x_2)$, whereas the second set to the transform plane $(\xi,\eta)$. Accordingly, the functions $(\beta,\gamma)$ entering the expressions for $u_j^*$ in (26) become

$$\beta=\rho,\quad \gamma=\left(a^2+\rho^2\right)^{1/2},\qquad \text{with } \rho\geq 0\ . \tag{27}$$

Now, in view of (22)-(24) and the newly introduced polar coordinates, we obtain

$$u_j(r,\theta,x_3=0) = \frac{1}{4\pi^2}\int_0^\infty\int_0^{2\pi} u_j^*(\rho,\varphi,x_3=0)\cdot e^{-i\rho r\cos(\varphi-\theta)}\rho\,d\rho\,d\varphi\ , \quad (j=1,2,3) \tag{28}$$

where the transformed displacements $u_j^*$ take now the following form

$$u_1^*(\rho,\varphi,x_3=0) = iF_1(\rho)\cos\varphi\ , \tag{29}$$

$$u_2^*(\rho,\varphi,x_3=0) = iF_1(\rho)\sin\varphi\ , \tag{30}$$

$$u_3^*(\rho,\varphi,x_3=0) = F_2(\rho)\ , \tag{31}$$

with

$$F_1(\rho) = -\frac{P(1-2\nu)}{2\mu\rho} + \frac{Pc\rho(1-2\nu)}{2\mu(1+c\rho^2)} + \frac{Pc(1-\nu)\rho\left[4c^2\gamma^2\rho(\gamma-\rho)-2c\rho^2-3+2\nu\right]}{2\mu(1+c\rho^2)N}\ , \tag{32a}$$

$$F_2(\rho) = \frac{P(1-\nu)}{\mu\rho} - \frac{P(1-\nu)c\rho}{\mu(1+c\rho^2)} - \frac{Pc(1-\nu)\rho\left[c(1+2c\rho^2-2c\rho\gamma)\gamma\rho-1+\nu\right]}{\mu(1+c\rho^2)N} . \tag{32b}$$

Further, in order to expose the circular symmetry of the Boussinesq problem, it is useful to resolve the displacement field in cylindrical polar coordinates. Thus, in view of (28)-(31), and making use of the well-known Bessel integral identity [70]

$$\int_0^{2\pi} e^{-i\rho r\cos(\varphi-\theta)} e^{in\varphi} d\varphi = 2\pi(-i)^n J_n(\rho r) \cdot e^{in\theta} , \quad n=0,1,2,3,..., \tag{33}$$

the displacement field appears in cylindrical polar coordinates as

$$u_r(r,\theta,x_3=0) = \frac{1}{2\pi}\int_0^\infty F_1(\rho)\rho J_1(\rho r)d\rho , \tag{34}$$

$$u_\theta(r,\theta,x_3=0) = 0 , \tag{35}$$

$$u_3(r,\theta,x_3=0) = \frac{1}{2\pi}\int_0^\infty F_2(\rho)\rho J_0(\rho r)d\rho . \tag{36}$$

where $J_n(\ )$ is the Bessel function of the first kind of order $n$. One may observe that the integrals in (34)-(36) are but inverse Hankel transforms. Moreover, it is noted that the tangential displacement $u_\theta$ vanishes while the radial and the vertical displacements are $\theta$-independent, as it is anticipated in view of the axisymmetry of the Boussinesq problem.

Next, each displacement component can be written in terms of three integrals, in the following form

$$u_r(r,x_3=0) = I_{class.} + I_{grad-1} + I_{grad-2} , \tag{37}$$

$$u_\theta(r,x_3=0) = 0 , \tag{38}$$

$$u_3(r,x_3=0) = II_{class.} + II_{grad-1} + II_{grad-2} , \tag{39}$$

where

$$I_{class.} = -\frac{P(1-2\nu)}{4\pi\mu}\int_0^\infty J_1(\rho r)d\rho \quad , \tag{40a}$$

$$I_{grad-1} = \frac{P(1-2\nu)}{4\pi\mu}\int_0^\infty \frac{c\rho}{(1+c\rho^2)}\rho J_1(\rho r)d\rho \quad , \tag{40b}$$

$$I_{grad-2} = \frac{P(1-\nu)}{4\pi\mu}\int_0^\infty \frac{c\rho\left[4c^2\gamma^2\rho(\gamma-\rho)-2c\rho^2-3+2\nu\right]}{(1+c\rho^2)N}\rho J_1(\rho r)d\rho \quad , \tag{40c}$$

and

$$II_{class.} = \frac{P(1-\nu)}{2\pi\mu}\int_0^\infty J_0(\rho r)d\rho \quad , \tag{41a}$$

$$II_{grad-1} = -\frac{P(1-\nu)}{2\pi\mu}\int_0^\infty \frac{c\rho}{(1+c\rho^2)}\rho J_0(\rho r)d\rho \quad , \tag{41b}$$

$$II_{grad-2} = -\frac{P(1-\nu)}{2\pi\mu}\int_0^\infty \frac{c\rho\left[c(1+2c\rho^2-2c\rho\gamma)\gamma\rho-1+\nu\right]}{(1+c\rho^2)N}\rho J_0(\rho r)d\rho \quad . \tag{41c}$$

The integrals in (40a), (40b), (41a) and (41b), can be obtained in closed form, while the integrals in (40c) and (41c) have to be evaluated numerically. Indeed, by employing standard results for Bessel integral identities and Hankel inverse transforms, we obtain [70,71]

$$I_{class.} = -\frac{P(1-2\nu)}{4\pi\mu}\frac{1}{r} \quad , \quad I_{grad-1} = \frac{P(1-2\nu)}{4\pi\mu}\frac{K_1(r/c^{1/2})}{c^{1/2}} \quad ,$$

$$II_{class.} = \frac{P(1-\nu)}{2\pi\mu}\frac{1}{r} \quad , \quad II_{grad-1} = -\frac{P(1-\nu)}{2\pi\mu}\frac{1}{r} + \frac{P(1-\nu)}{4\mu c^{1/2}}\left[I_0(r/c^{1/2}) - L_0(r/c^{1/2})\right] \quad , \tag{42a-d}$$

where $I_n(\ )$ and $K_n(\ )$ are the $n^{th}$-order modified Bessel functions of the first and second kind, respectively, and $L_n(\ )$ is the $n^{th}$-order modified Struve function [72].

Finally, in the limit case of classical linear elasticity ($c \to 0$), the second and third terms in the RHS of each one of equations (37)-(39) vanish and the surface displacements then become

$$u_r^{class.}(r, x_3 = 0) = -\frac{P(1-2\nu)}{4\pi\mu}\frac{1}{r}, \tag{43}$$

$$u_\theta^{class.}(r, x_3 = 0) = 0, \tag{44}$$

$$u_3^{class.}(r, x_3 = 0) = \frac{P(1-\nu)}{2\pi\mu}\frac{1}{r}, \tag{45}$$

which is the classical Boussinesq field that is *unbounded* at the point of application of the concentrated load [1-5].

## 5  Solution and numerical results

From equations (37)-(42), the final results for the surface displacements may follow. We provide them below in normalized form

$$\hat{u}_r(r', x_3 = 0) = (1-2\nu)\left[-\frac{1}{r'} + K_1(r')\right] + (1-\nu)\int_0^\infty G(\rho')\,\rho' J_1(\rho' r')d\rho', \tag{46}$$

$$\hat{u}_3(r', x_3 = 0) = \frac{\pi(1-\nu)}{2}\left[I_0(r') - L_0(r')\right] - (1-\nu)\int_0^\infty H(\rho')\,\rho' J_0(\rho' r')d\rho', \tag{47}$$

with

$$G(\rho') = \frac{\rho'\left[4\gamma'^2\rho'(\gamma'-\rho') - 2\rho'^2 - 3 + 2\nu\right]}{(1+\rho'^2)\Lambda(\rho')}, \tag{48a}$$

$$H(\rho') = \frac{\rho'\left[\left(1+2\rho'^2-2\gamma'\rho'\right)\gamma'\rho'-1+v\right]}{(1+\rho'^2)\Lambda(\rho')}, \tag{48b}$$

where the superposed caret in the displacements denotes normalized quantities that will be defined below. Also, $\rho' = c^{1/2}\rho$ and $r' = c^{-1/2}r$ denote normalized distances (dimensionless), and $\gamma' = (1+\rho'^2)^{1/2}$. The function $\Lambda(\rho')$ is then given by

$$\Lambda(\rho') = 4(\rho'\gamma')^3 - (1+2\rho'^2)\left(2(\rho'\gamma')^2 - v + 1\right). \tag{49}$$

In addition, the normalized displacements are dimensionless quantities defined as follows

$$\hat{u}_r = \frac{4\pi\mu c^{1/2}}{P}u_r, \quad \hat{u}_3 = \frac{2\pi\mu c^{1/2}}{P}u_3. \tag{50}$$

Now, examining the asymptotic behavior of the solution (to determine the possibility of singularities), we note that as $r' \to 0$ the following asymptotic relations hold [71,72]

$$K_1(r') = \frac{1}{r'} + O(r'), \quad L_0(r') = \frac{2r'}{\pi} + O(r'^3), \quad I_0(r') = 1 + O(r'^2). \tag{51}$$

Moreover, taking into account that: (i) $G(\rho')$ and $H(\rho')$ in (48) are continuous functions in $\rho' \in [0,\infty)$, (ii) $|G(\rho')\rho' J_1(\rho'r')| \leq G(\rho')\rho'$ and $|H(\rho')\rho' J_0(\rho'r')| \leq H(\rho')\rho'$ for $(r',\rho') \in [0,\infty) \times [0,\infty)$, and (iii) $\int_0^\infty G(\rho')\rho'd\rho' < \infty$ and $\int_0^\infty H(\rho')\rho'd\rho' < \infty$. It can readily be shown, employing the Weierstrass M-test [73], that both integrals in (46) and (47) are *uniformly convergent* in $r' \in [0,\infty)$, and thus *regular* as $r' \to 0$. The latter integrals are evaluated numerically employing MATHEMATICA$^{\text{TM}}$ algorithms that take into account their oscillatory character.

In light of the above, one can immediately infer that the undesirable Cauchy-type singularities predicted by the classical theory for the displacements in Boussinesq problem are eliminated within the context of this simple version of dipolar gradient elasticity involving only one additional material constant. The occurrence of a *bounded* displacement field has certain implications in the behavior of the strain energy density. Indeed, by employing the general expressions for the transformed displacements (22)-(24) in conjunction with the inverse Laplace transform in (26) for $x_3 \neq 0$, we deduce (with the aid of asymptotic analysis) that the strains and the gradient of strains behave at most as

$$\varepsilon_{pq} = O(\ln R) \quad \text{and} \quad \partial_k \varepsilon_{pq} = O(R^{-1}) \quad \text{as} \quad R \to 0, \tag{52a,b}$$

where $R$ is the radial distance in spherical coordinates defined as $R = (x_1 + x_2 + x_3)^{1/2}$. Thus, in our case, in view of the above result and equation (8), we derive that the strain-energy density $W$ behaves at most as $W \approx (\partial_k \varepsilon_{pq})^2 = O(R^{-2})$. It can further be checked that the (total) strain energy $U$ in a small hemispherical volume around the point of application of the concentrated load can be written in the form: $U = C \int_0^{R_0} W R^2 dR$, where $C$ is a constant that depends upon the elastic material constants and the variation of the displacements with the spherical angular coordinates, and $R_0$ is the radius of the small hemisphere. Consequently, the (total) strain energy $U$ is *bounded* for $R_0 > 0$, which, in turn, implies that the Kirchhoff-type theorem established in [26] for gradient elasticity guarantees solution uniqueness for the present problem.

This finding is in marked contrast with the cases of the Boussinesq problem in the theories of classical elasticity [2,4], and micropolar and couple-stress elasticity [49-52]. In the latter cases, the strain energy $U$ becomes unbounded near the point of application of the load. Indeed, in classical elasticity the strains exhibit an $O(R^{-2})$ behavior, whereas in micropolar and couple-stress elasticity we have $\varepsilon_{pq} = O(R^{-2})$ and $\partial_p \omega_q = O(R^{-1})$, hence the strain energy density behaves at most as $W \approx (\varepsilon_{pq})^2 = O(R^{-4})$, thus leading to an unbounded strain energy. Clearly, in these cases, a simple Kirchhoff-type theorem for uniqueness does not suffice – an augmented theorem should be established. As Sternberg and Eubanks [74], and Hartranft and Sih [75] indicated, this can be done by following a limiting process and by imposing certain restrictions on the behavior of the solution near the singular point.

We emphasize that in our case the simple Kirchhoff-type theorem of gradient elasticity [26] suffices to guarantee uniqueness of the solution.

Some numerical results are now presented. The integrals in (46) and (47) are evaluated numerically by employing MATHEMATICA$^{TM}$. Figure 5 depicts the variation of the normalized vertical surface displacement $\hat{u}_3(r',0)$ with the normalized distance $r'$ from the point of application of the concentrated load. It is clearly observed that the gradient solution exhibits a *bounded* vertical displacement at the point of application of the load. Thus, elimination of the undesirable Cauchy-type singularity of the classical Boussinesq solution is achieved within gradient elasticity theory. This finding is also in agreement with the numerical result provided by Karlis et al. [47] using the boundary element method. It should remarked that other gradient type theories like the couple-stress elasticity and the micropolar (Cosserat) elasticity are not able of eliminating the singularity in the displacement field in the 3D Boussinesq problem. In fact, in these theories the Cauchy-type singularity is retained just as in the classical theory, although the detailed structure of the displacement field is altered [49-52]. In addition, Figure 6 displays the variation of the maximum (bounded) vertical surface displacement $\hat{u}_3(0,0)$ under the point of application of the concentrated load with respect to Poisson's ratio $\nu$. As anticipated on intuitive grounds, the maximum deflection decreases indeed with increasing values of Poisson's ratio.

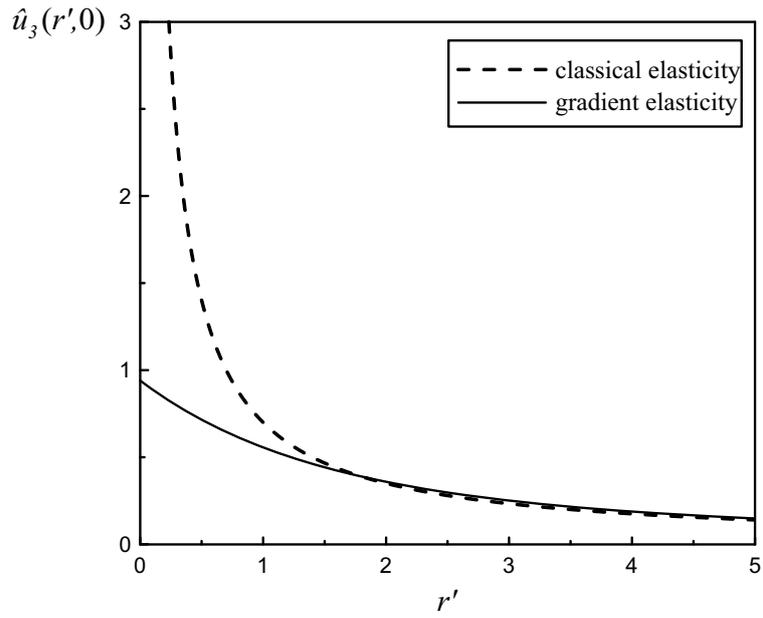

**Fig. 5** Normalized vertical surface displacement due to a concentrated normal load as provided by the gradient and classical theories of elasticity. The Poisson's ratio is $\nu = 0.3$.

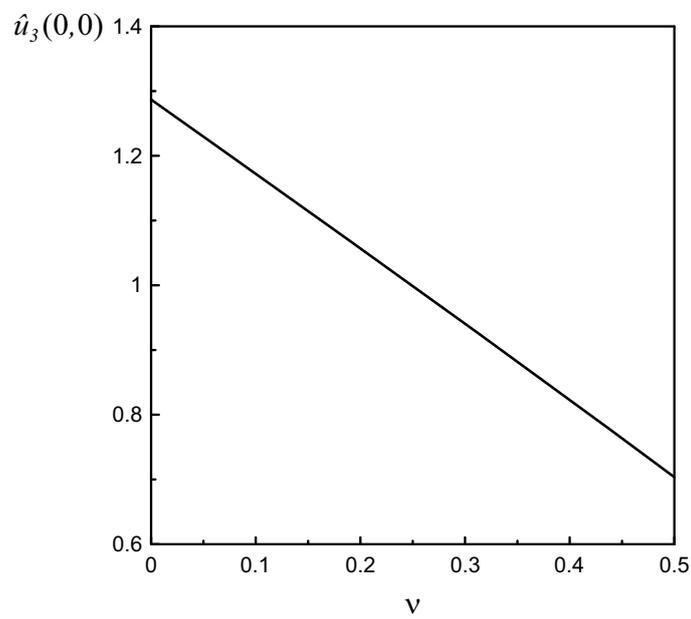

**Fig. 6** Variation of the maximum surface vertical displacement under the point of application of the load versus the Poisson's ratio $\nu$ in gradient elasticity.

Based on our numerical results and bearing in mind Eq. (50), a very good estimate for the maximum vertical surface displacement, depicted in Figure 6, is given by the expression $u_3(0,0) = \left(P/2\pi c^{1/2}\mu\right)(1.286 - 1.166\nu)$. It is worth noticing that the latter expression can find practical application in problems of Soil Mechanics and Tribology.

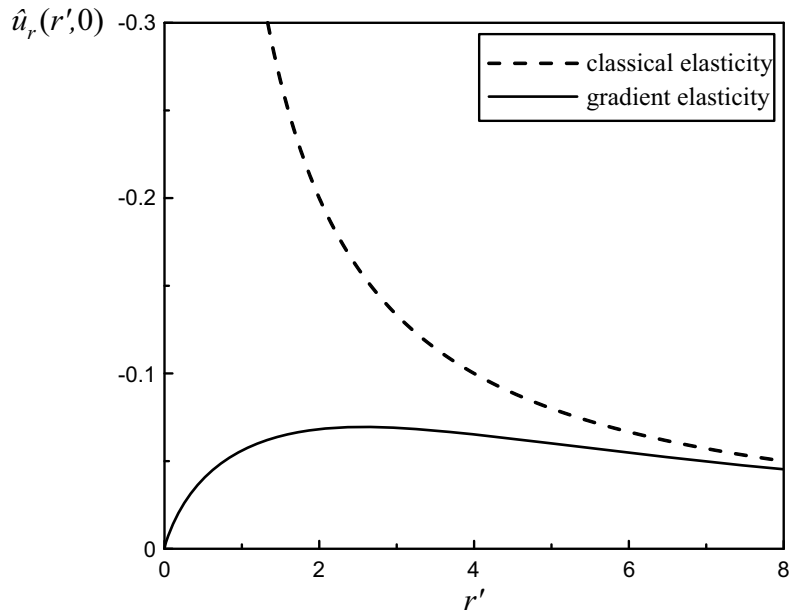

**Fig.7** Normalized radial surface displacement due to a concentrated normal load as provided by the gradient and classical theories of elasticity. The Poisson's ratio is $\nu = 0.3$.

Figure 7 depicts the variation of the normalized radial surface displacement $\hat{u}_r(r',0)$ with the distance from the point of application of the concentrated load. It is observed that gradient elasticity predicts zero radial surface displacement at the point of application of the load. Such a behavior seems to be more natural, taking into account the axisymmetry of the Boussinesq problem, as compared to the singular behavior present in the classical solution. Finally, as shown in Figure 8, the radial displacement exhibits a bounded maximum in a small zone around the concentrated load ($r \approx 2c^{1/2}$), which strongly depends on the Poisson's ratio.

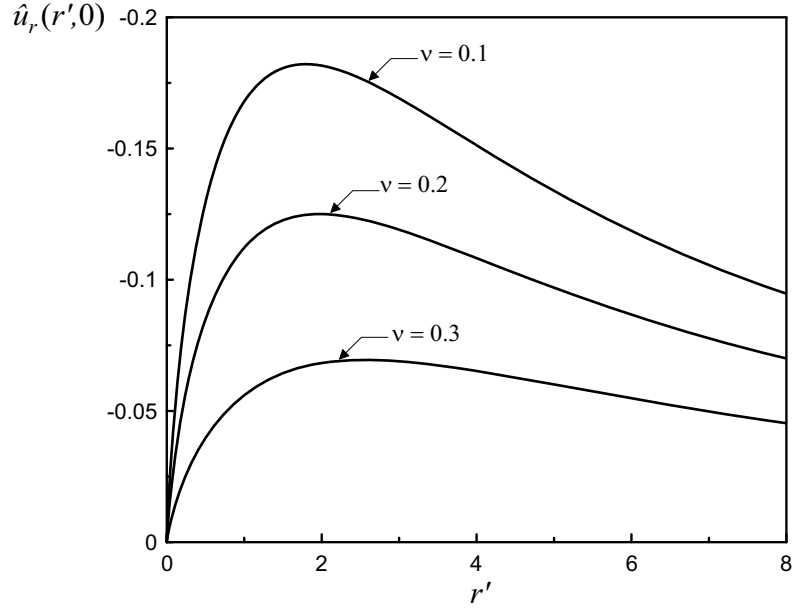

**Fig. 8** Normalized radial surface displacement in gradient elasticity for various Poisson's ratios.

## 6 Discussion and concluding remarks

In the present work, we examined the 3D axisymmetric Boussinesq problem in a body with microstructural properties governed by dipolar gradient elasticity. Our main concern was to determine possible deviations from the predictions of classical linear elastostatics when a more refined theory is employed to attack this problem. A simple but yet rigorous version of dipolar gradient elasticity was employed by considering an isotropic linear expression of the elastic strain-energy density that involves only three material constants (the two Lamé constants and the so-called gradient coefficient). In this way, this theory can be viewed as a first-step extension of classical elasticity.

The solution method is based on integral transforms and is exact. Our results show significant departure from the predictions of classical elasticity. Indeed, bounded and continuous displacements are predicted at the point of application of the concentrated load. In particular, the vertical displacement attains a constant (bounded) value at the point of application of the load. This displacement depends upon the load intensity and the material constants, thus providing a sound meaning of the load-bearing capacity of the medium. In addition, the radial displacement becomes zero at the point of application of the load, a finding that can be justified by the axisymmetric character of the Boussinesq problem. The behavior of the displacement field now seems more natural than the singular behavior exhibited in the classical solution. Finally, we note that analogous results were obtained in the 2D Flamant-Boussinesq and Kelvin problems in dipolar gradient elasticity (Georgiadis and Anagnostou [14]).

We also remark that our solution may serve as a Green's function for more general loadings. In addition, the present results can be used to accomplish integral-equation solutions for 3D contact problems in dipolar gradient elasticity.

**Acknowledgement**

The authors acknowledge with thanks support from the 'ΠΕΒΕ 2010' programme of NTU Athens (# 65/1843, title of the individual project: 'Dynamic Fracture and Contact within the Framework of Generalized Continua and Thermo-Mechanics').

**Appendix A**

The displacement equations of equilibrium (10b) can be written in the following form

$$\left(1-c\nabla^2\right)\left[\nabla^2 u_1 + \left(1-2\nu\right)^{-1}\partial_1 e\right] = 0 , \tag{A1a}$$

$$\left(1-c\nabla^2\right)\left[\nabla^2 u_2 + \left(1-2\nu\right)^{-1}\partial_2 e\right] = 0 , \tag{A1b}$$

$$\left(1-c\nabla^2\right)\left[\nabla^2 u_3 + \left(1-2\nu\right)^{-1}\partial_3 e\right] = 0 , \tag{A1c}$$

where $e \equiv \nabla \cdot \mathbf{u}$ is the dilatation. Applying now the bilateral Laplace transform (14a) to the above coupled system of PDEs and noting that the bilateral Laplace transform of any derivative $\partial/\partial x_1^n$ ($\partial/\partial x_2^n$) ($n \in N^*$) w.r.t. $x_1$ ($x_2$) yields in the transform domain an image $p^n$ ($q^n$), we obtain readily the following system of ODEs

$$\left(1 - cp^2 - cq^2 - c\frac{d^2}{dx_3^2}\right)\left[\left(p^2 + q^2 + \frac{d^2}{dx_3^2}\right)u_1^* + \left(1-2\nu\right)^{-1} p e^*\right] = 0 , \tag{A2a}$$

$$\left(1 - cp^2 - cq^2 - c\frac{d^2}{dx_3^2}\right)\left[\left(p^2 + q^2 + \frac{d^2}{dx_3^2}\right)u_2^* + (1-2\nu)^{-1} qe^*\right] = 0, \tag{A2b}$$

$$\left(1 - cp^2 - cq^2 - c\frac{d^2}{dx_3^2}\right)\left[\left(p^2 + q^2 + \frac{d^2}{dx_3^2}\right)u_3^* + (1-2\nu)^{-1} \frac{de^*}{dx_3}\right] = 0, \tag{A2c}$$

where $e^* = pu_1^* + qu_2^* + \dfrac{du_3^*}{dx_3}$. The above system can now be written in a compact form as

$$[K]\begin{bmatrix} u_1^* \\ u_2^* \\ u_3^* \end{bmatrix} \equiv \begin{bmatrix} 0 \\ 0 \\ 0 \end{bmatrix}, \tag{A3}$$

where the coefficients of the differential operator $[K]$ are provided in Eqs. (16).

**Appendix B**

Here, the transformed general expressions for the stresses and dipolar stresses that enter the boundary conditions in (12) and (13) are provided. These are obtained by applying the transform (14a) to the constitutive equations (9), in conjunction with (18)-(21), as

$$\tau_{11}^*(p,q,x_3) = \lambda\left(qu_2^* + \frac{du_3^*}{dx_3}\right) + (\lambda + 2\mu)pu_1^*$$

$$= 2\mu\left[A_1\beta - A_2q - A_3\left(3 - 2\nu + \frac{p^2}{\beta}x_3\right)\right]e^{-\beta x_3}$$

$$+ \frac{2\mu}{1-2\nu}\left[B_1(1-\nu)p + B_2\nu q - B_3\nu\gamma\right]e^{-\gamma x_3}, \tag{B1}$$

$$\tau_{22}^*(p,q,x_3) = \lambda\left(pu_1^* + \frac{du_3^*}{dx_3}\right) + (\lambda + 2\mu)qu_2^*$$

$$= 2\mu\left[A_2q - A_3\left(2\nu + \frac{q^2}{\beta}x_3\right)\right]e^{-\beta x_3} + \frac{2\mu}{1-2\nu}\left[B_1\nu p + B_2(1-\nu)q - B_3\nu\gamma\right]e^{-\gamma x_3},$$

$$\tau_{12}^*(p,q,x_3) = \mu\left(qu_1^* + pu_2^*\right)$$

$$= \mu p^{-1}\left[A_1 q\beta + A_2\left(p^2 - q^2\right) - A_3\beta^{-1}\left((3-4\nu)q\beta + 2p^2 q x_3\right)\right]e^{-\beta x_3}$$

$$+ \mu\left[B_1 q + B_2 p\right]e^{-\gamma x_3} \;, \tag{B3}$$

$$\tau_{33}^*(p,q,x_3) = \lambda\left(pu_1^* + qu_2^*\right) + (\lambda + 2\mu)\frac{du_3^*}{dx_3}$$

$$= 2\mu\left[-A_1\beta + A_3(1-2\nu-\beta x_3)\right]e^{-\beta x_3} + \frac{2\mu}{1-2\nu}\left[B_1\nu p + B_2\nu q - B_3(1-\nu)\gamma\right]e^{-\gamma x_3} \;,$$

$$\tag{B4}$$

$$\tau_{31}^*(p,q,x_3) = \mu\left(pu_3^* + \frac{du_1^*}{dx_3}\right)$$

$$= \mu p^{-1}\left[A_1\left(2p^2 + q^2\right) + A_2 q\beta - A_3\beta^{-1}\left(4(1-\nu)p^2 + (3-4\nu)q^2 - 2p^2\beta x_3\right)\right]e^{-\beta x_3}$$

$$- \mu\left[B_1\gamma - B_3 p\right]e^{-\gamma x_3} \;, \tag{B5}$$

$$\tau_{32}^*(p,q,x_3) = \mu\left(qu_3^* + \frac{du_2^*}{dx_3}\right)$$

$$= \mu\left[A_1 q - A_2\beta - A_3 q\beta^{-1}(1-2\beta x_3)\right]e^{-\beta x_3} - \mu\left[B_2\gamma - B_3 q\right]e^{-\gamma x_3} \;, \tag{B6}$$

$$m_{13k}^*(p,q,x_3) = cp\tau_{3k}^* \;, \qquad m_{23k}^*(p,q,x_3) = cq\tau_{3k}^* \;, \qquad m_{3kl}^*(p,q,x_3) = c\frac{d\tau_{kl}^*}{dx_3} \;, \tag{B7}$$

where the Latin indices $(k,l)$ span the range $(1,2,3)$, and $\lambda = 2\mu\nu/(1-2\nu)$.

**Appendix C**

Transforming the boundary conditions (12)-(13) with (14a) and enforcing (B1)-(B7) provides a $6\times 6$ system of algebraic equations for the unknown functions $(A_j, B_j)$, with $(j=1,2,3)$. This system is solved using manipulations both by hand and the symbolic program MATHEMATICA$^{\text{TM}}$. The quantities $(A_j, B_j)$ are functions of $(p, q)$ and depend upon the material parameters

$$A_1 = \frac{P}{\mu \beta N} \left[ (3 - 2\nu) c^2 \beta^3 \gamma - (1 - \nu)(2c^2 \beta^2 \gamma^2 - \nu + 1) \right], \tag{C1}$$

$$A_2 = -\frac{Pq}{2\mu \beta^2 N} \left[ 4\nu c^2 \beta^3 \gamma + (1 - 2\nu)(2c^2 \beta^2 \gamma^2 - \nu + 1) \right], \tag{C2}$$

$$A_3 = \frac{P}{2\mu N} \left[ 2c^2 \beta^2 \gamma (\beta - \gamma) - 1 + \nu \right], \tag{C3}$$

$$B_1 = \frac{Pcp}{\mu N} \left[ c\gamma^2 - \nu \right], \tag{C4}$$

$$B_2 = \frac{Pcq}{\mu N} \left[ c\gamma^2 - \nu \right], \tag{C5}$$

$$B_3 = -\frac{Pc\beta^2}{\mu \gamma N} \left[ c\beta^2 + \nu \right], \tag{C6}$$

with

$$N = 4(c\beta\gamma)^3 - (1 + 2c\beta^2)(2(c\beta\gamma)^2 - \nu + 1). \tag{C7}$$

**References**

1. Love, A.E.H.: A Treatise on the Mathematical Theory of Elasticity. Cambridge University Press, New York (1952)
2. Timoshenko, S.P., Goodier, J.N.: Theory of Elasticity. McGraw-Hill, New York (1970)
3. Fung, Y.C.: Foundations of Solid Mechanics. Prentice-Hall, Englewood Cliffs, NJ (1965)
4. Barber, J.R.: Elasticity. Kluwer Academic Publishers, Dordrecht (2002)
5. Selvadurai, A.P.S.: On Boussinesq's problem. Int. J. Eng. Sci. **39**, 317-322 (2001)
6. Bowles, J.E.: Foundation Analysis and Design. Mcgraw-Hill, New York, (1996).
7. Davis, R.O., Selvadurai, A.P.S.: Elasticity and Geomechanics. Cambridge University Press, Cambridge (1996).

8. Gladwell, G.M.L.: Contact Problems in the Classical Theory of Elasticity. Kluwer Academic Publishers, Alphen aan den Rijn (1980)
9. Hills, D.A., Nowell, D., Sackfield, A.: Mechanics of Elastic Contacts. Butterworth-Heinemann, Oxford (1993)
10. Beskos, D.E. (Ed.): Boundary Element Methods in Mechanics. North-Holland, Amsterdam (1987)
11. Toupin, R.A.: Elastic materials with couple-stresses. Arch. Rational Mech. Anal. **11**, 385-414 (1962)
12. Mindlin, R.D.: Micro-structure in linear elasticity. Arch. Rational Mech. Anal. 16, 51-78 (1964)
13. Georgiadis, H.G., Vardoulakis, I., Velgaki, E.G.: Dispersive Rayleigh-wave propagation in microstructured solids characterized by dipolar gradient elasticity. J. Elast. **74**, 17-45 (2004)
14. Georgiadis, H.G., Anagnostou, D.S.: Problems of the Flamant-Boussinesq and Kelvin type in dipolar gradient elasticity. J. Elast. **90**, 71-98 (2008)
15. Neff, P.: Existence of minimizers in nonlinear elastostatics of micromorphic solids. Encyclopedia of Thermal Stresses, Springer (2013)
16. Ghiba, I.D., Neff, P., Madeo, A., Placidi, L., Rosi, G.: The relaxed linear micromorphic continuum: existence, uniqueness and continuous dependence in dynamics. arXiv:1308.3762.
17. Neff, P., Ghiba, I.D., Madeo, A., Placidi, L., Rosi, G.: A unifying perspective: the relaxed linear micromorphic continuum. arXiv:1308.3219.
18. Maugin, G.A.: Mechanics of generalized continua: What do we mean by that? In: Maugin, G.A. Metrikine A.V. (Eds.), Mechanics of Generalized Continua. One Hundred Years After the Cosserats. Advances in Mathematics and Mechanics **21**, Springer, New York, 253-262 (2010)
19. Cook, T.S., Weitsman, Y.: Strain gradient effects around spherical inclusions and cavities. Int. J. Solids Struct. **2**, 393-406 (1966)
20. Eshel, N.N., Rosenfeld, G.: Effects of strain-gradient on the stress-concentration at a cylindrical hole in a field of uniaxial tension. J. Eng. Math. **4**, 97-111 (1970)
21. Eshel, N.N., Rosenfeld, G.: Axisymmetric problems in elastic materials of grade two. J. Franklin Inst. 299, 43-51 (1975)
22. Chen, J.Y., Huang, Y., Ortiz, M.: Fracture analysis of cellular materials: A strain gradient model. J. Mech. Phys. Solids 46, 789-828 (1998)
23. Shi, M.X., Huang, Y., Hwang, K.C.: Fracture in the higher-order elastic continuum. J. Mech. Phys. Solids 48, 2513-2538 (2000)

24. Georgiadis, H.G.: The mode-III crack problem in microstructured solids governed by dipolar gradient elasticity: Static and dynamic analysis. ASME J. Appl. Mech. **70**, 517-530 (2003)
25. Radi, E., Gei, M.: Mode III crack growth in linear hardening materials with strain gradient effects. Int. J. Fract. **130**, 765-785 (2004)
26. Grentzelou, C.G., Georgiadis, H.G.: Uniqueness for plane crack problems in dipolar gradient elasticity and in couple-stress elasticity. Int. J. Solids Struct. **42**, 6226-6244 (2005)
27. Grentzelou, C.G., Georgiadis, H.G.: Balance laws and energy release rates for cracks in dipolar gradient elasticity. Int. J. Solids Struct. **45**, 551-567 (2008)
28. Gourgiotis, P.A., Georgiadis, H.G.: Plane-strain crack problems in microstructured solids governed by dipolar gradient elasticity. J. Mech. Phys. Solids **57**, 1898-1920 (2009)
29. Aravas, N., Giannakopoulos, A.E.: Plane asymptotic crack-tip solutions in gradient elasticity. Int. J. Solids Struct. **46**, 4478-4503 (2009)
30. Aslan, O., Forest, S.: Crack growth modelling in single crystals based on higher order continua. Comp. Mater. Sci. **45**, 756-761 (2009)
31. Gourgiotis, P.A., Sifnaiou, M.D., Georgiadis, H.G.: The problem of sharp notch in microstructured solids governed by dipolar gradient elasticity. Int. J. Fract. **166**, 179-201 (2010)
32. Piccolroaz, A., Mishuris, G., Radi, E.: Mode III interfacial crack in the presence of couple stress elastic materials. Eng. Fract. Mech. **80**, 60-71 (2012)
33. Sciarra, G., Vidoli, S.: Asymptotic fracture modes in strain-gradient elasticity: Size effects and characteristic lengths for isotropic materials. J. Elast. **113**, 27-53 (2013)
34. Maugin, G.A., Miled, A.: Solitary waves in micropolar elastic crystals, Int. J. Eng. Sci. **24**, 1477-1499 (1986)
35. Engelbrecht, J. Berezovski, A., Pastrone, F., Braun, M.: Waves in microstructured materials and dispersion, Phil. Mag. **85**, 4127-4141 (2005)
36. dell'Isola, F., Madeo, A., Placidi, L.: Linear plane wave propagation and normal transmission and reflection at discontinuity surfaces in second gradient 3D Continua. ZAMM **92**, 52-71 (2012)
37. Polyzos, D., Fotiadis, D.I.: Derivation of Mindlin's first and second strain gradient elastic theory via simple lattice and continuum models. Int. J. Solids Struct. **49**, 470-480 (2012)
38. Gourgiotis, P.A., Georgiadis, H.G., Neocleous, I.: On the reflection of waves in half-spaces of microstructured materials governed by dipolar gradient elasticity. Wave Motion **50**, 437-455 (2013)

39. Madeo, A., dell'Isola, F., Ianiro, N., Sciarra, G.: A Variational deduction of second gradient poroelasticity II: An application to the consolidation problem. J. Mech. Mater. Struct. **3**, 607-625 (2008).

40. Lazar, M., Maugin, G.A.: Defects in gradient micropolar elasticity-II: edge dislocation and wedge disclination. J. Mech. Phys. Solids **52**, 2285-2307 (2004)

41. Lazar, M., Maugin, G.A.: Nonsingular stress and strain fields of dislocations and disclinations in first strain gradient elasticity. Int. J. Eng. Sci. **43**, 1157-1184 (2005)

42. Forest, S.: Some links between Cosserat, strain gradient crystal plasticity and the statistical theory of dislocations. Philos. Mag. A **88**, 3549-3563 (2008).

43. Exadaktylos, G: Some basic half-plane problems of the cohesive elasticity theory with surface energy. Acta Mechanica **133**, 175-198 (1999)

44. Lazar, M., Maugin, G.A.: A note on line forces in gradient elasticity. Mech. Res. Commun. **33**, 674-680 (2006)

45. White, R.M.: Surface elastic waves. Proceedings IEEE 58, 1238-1276 (1970)

46. Farnell, G.W.: Types and properties of surface waves. In: Oliner, A.A. (Ed.), Acoustic Surface Waves. Springer-Verlag, Berlin, 13-60 (1978)

47. Karlis, G.F., Tsinopoulos, S.V., Polyzos, D.: Boundary element analysis of gradient elastic problems. In: Manolis, G.D., Polyzos, D. (Eds.) Recent Advances in Boundary Element Methods: A Volume to honor Professor D.E. Beskos. Springer, New-York, 239-252 (2009)

48. Gao, X.L., Zhou, S.S.: Strain gradient solutions of half-space and half-plane contact problems. ZAMP. DOI 10.1007/s00033-012-0273-1 (2012)

49. Dhaliwal, R.S.: The axisymmetric Boussinesq problem in micropolar theory of elasticity. Arch. Mech. **24**, 645-653 (1972)

50. Sandru, N.: The influence of couple-stresses on singular stress concentrations in micropolar elastic half-space. Int. J. Eng. Sci. 13, 631-639 (1975)

51. Khan, S.M., Dhaliwal, R.S.: Axisymmetric problem for a half-space in micropolar theory of elasticity. J. Elast. 7, 13-32 (1977)

52. Dyszlewicz, J.: Generalized Papkovich-Neuber's representation in the micropolar elasticity. The Boussinesq-type problem for a micropolar half-space. Bull. Pol. Acad. Tech. 42, 1-12 (1994)

53. Nowinski, J.L.: On the three-dimensional Boussinesq for an elastic nonlocal medium. Int. J. Eng. Sci. 28, 1245-1251 (1990)

54. Simmonds, J.G., Warne, P.G.: Notes on the nonlinearly elastic Boussinesq problem. J. Elast. 34, 69-82 (1994)

55. Georgiadis, H.G., Grentzelou, C.G.: Energy theorems and the J-integral in dipolar gradient elasticity. Int. J. Solids Struct. **43**, 5690-5712 (2006)

56. Bleustein, J.L.: A note on the boundary conditions of Toupin's strain-gradient theory. Int. J. Solids Struct. **3**, 1053-1057 (1967)

57. Mindlin, R.D., Eshel, N.N.: On first-gradient theories in linear elasticity. Int. J. Solids Struct. **4**, 109-124 (1968)

58. Green, A.E., Rivlin, R.S.: Multipolar continuum mechanics. Arch. Ration. Mech. Anal. **17**, 113-147 (1964).

59. Jaunzemis, W.: Continuum Mechanics. McMillan, New York (1967)

60. Alibert, J., Seppecher, P., dell'Isola, F.: Truss modular beams with deformation energy depending on higher displacement gradients. Math. Mech. Solids **8**, 51-73 (2003).

61. dell'Isola, F., Seppecher, P.: The relationship between edge contact forces, double forces and interstitial working allowed by the principle of virtual power. C.R. Acad. Sci. IIB **321**, 303-308 (1995).

62. dell'Isola, F., Seppecher, P., Madeo, A.: How contact interactions may depend on the shape of Cauchy cuts in N-*th* gradient continua: approach "`a la D'Alembert". ZAMP **63**, 1119-1141 (2012)

63. dell'Isola, F., Sciarra, G., Vidoli, S.: Generalized Hooke's law for isotropic second gradient materials. Proc. Roc. S. A **465**, 2177-2196 (2009)

64. Gourgiotis, P.A., Georgiadis, H.G., Petrou, Z.P.: Solution of two-dimensional concentrated load problems in the exact Toupin-Mindlin theory of dipolar gradient elasticity, submitted. (2013)

65. van der Pol, B., Bremmer, H.: Operational Calculus Based on the Two-Sided Laplace Integral. Cambridge University Press, Cambridge (1950)

66. Carrier, G.A., Krook, M., Pearson, C.E.: Functions of a Complex Variable. McGraw-Hill, New York (1966)

67. Brock, L.M., Rodgers, M.J.: Steady-state response of a thermoelastic half-space to the rapid motion of surface thermal/mechanical loads. J. Elast. 47, 225-240 (1997)

68. Georgiadis, H.G., Lykotrafitis, G.: Rayleigh waves generated by a thermal source: A three dimensional transient solution. J. App. Mech. **72**, 129-138 (2005)

69. Sneddon, I.N.: Fourier Transforms. McGraw-Hill, New York (1951)

70. Watson, G.N.: The Theory of Bessel Functions. Cambridge University Press, Cambridge (1966)

71. Gradshteyn, I.S., Ryzhik, I.M.: Table of Integrals, Series and Products. Academic Press, New York (1980)

72. Abramowitz, M., Stegun, I.A.: Handbook of Mathematical Functions. Dover, New York (1982)
73. Widder, D.V.: Advanced Calculus. Prentice-Hall, New York (1961)
74. Sternberg, E., Eubanks, R.A.: On the concept of concentrated loads and an extension of the uniqueness theorem in the linear theory of elasticity. J. Rational Mech. Anal. **4**, 135-168 (1955)
75. Hartranft, R.J., Sih, G.C.: Uniqueness of the concentrated-load problem in the linear theory of couple-stress elasticity. Meccanica **3**, 195-198 (1969)